\documentclass[lettersize,journal,compsoc]{IEEEtran}
\usepackage{amsmath,amsfonts}
\usepackage{algorithmic}
\usepackage[ruled,vlined]{algorithm2e}
\usepackage{array}
\usepackage[caption=false,font=normalsize,labelfont=sf,textfont=sf]{subfig}
\usepackage{textcomp}
\usepackage{stfloats}
\usepackage{url}
\usepackage{verbatim}
\usepackage{graphicx}
\usepackage{cite}
\usepackage{tikz}
\usepackage{jabbrv}
\usepackage{amssymb}   
\usepackage{pifont}    
\usepackage{graphicx}
\usepackage{booktabs}
\usepackage{multirow}
\usepackage{xcolor}
\usepackage{tabularx}
\usepackage{makecell}
\usepackage{float}
\usepackage{placeins}
\usepackage{ragged2e}
\usepackage{subcaption}

\newcommand{\circnum}[1]{%
  \tikz[baseline=(char.base)]{
    \node[
      shape=circle,
      draw=black,
      fill=black,
      inner sep=0pt,
      minimum size=10pt
    ] (char) {\textcolor{white}{\scriptsize #1}};
  }%
}

\newcolumntype{?}{!{\vrule width 1pt}}

\begin{document}

\title{PDRIMA: A Policy-Driven Runtime Integrity Measurement and Attestation Approach for ARM TrustZone-based TEE}

\author{Jingkai Mao, Xiaolin Chang 
\IEEEcompsocitemizethanks{
\IEEEcompsocthanksitem Jingkai Mao, and Xiaolin Chang are with the Beijing Key Laboratory of Security and Privacy in Intelligent Transportation, Beijing Jiaotong University, P.R.China. (e-mail: \{23111143, xlchang\}@bjtu.edu.cn)
}}

\markboth{Journal of \LaTeX\ Class Files,~Vol.~14, No.~8, August~2021}%
{Shell \MakeLowercase{\textit{et al.}}: A Sample Article Using IEEEtran.cls for IEEE Journals}

\IEEEpubid{}

\IEEEtitleabstractindextext{
\begin{abstract}
\justifying
Trusted Execution Environments (TEEs) such as ARM TrustZone are widely used in IoT and embedded devices to protect sensitive code and data. However, most existing defenses focus on secure boot or REE-side monitoring and provide little visibility into the runtime integrity of the TEE. This leaves TrustZone-based devices exposed to persistent TEE compromises.

We propose Policy-Driven Runtime Integrity Measurement and Attestation (PDRIMA), a runtime integrity protection approach for TrustZone-based TEEs. PDRIMA systematically analyzes TEE attack surfaces and introduces two in-TEE subsystems: a Secure Monitor Agent (SMA) that performs policy-driven measurement, appraisal, logging, and time-based re-measurement over the TEE kernel, static components, user-TAs, and security-critical system calls; and a Remote Attestation Agent (RAA) that aggregates tamper-evident evidence and exposes a remote attestation protocol for verifying. We analyze PDRIMA’s security against identified attack surfaces, implement a prototype on OP-TEE for Raspberry Pi 3B+, and evaluate its performance overhead to indicate its practicability.
\end{abstract}

}

\maketitle
\section{Introduction}

\IEEEPARstart{T}{he} Internet of Things (IoT) and embedded systems are now widely used in daily life, powering intelligent transportation, unmanned aerial vehicles, smart homes. However, large-scale deployments of heterogeneous, resource-constrained devices expose many security vulnerabilities that can be exploited by attackers~\cite{sun2024survey}. Although numerous defenses have been proposed, they are still difficult to deploy consistently across diverse hardware and often fail to withstand sophisticated attacks~\cite{zhao2024trusted}.

\begin{figure}[!t]
\includegraphics[width=3.5in,trim=15 37 30 18,clip]{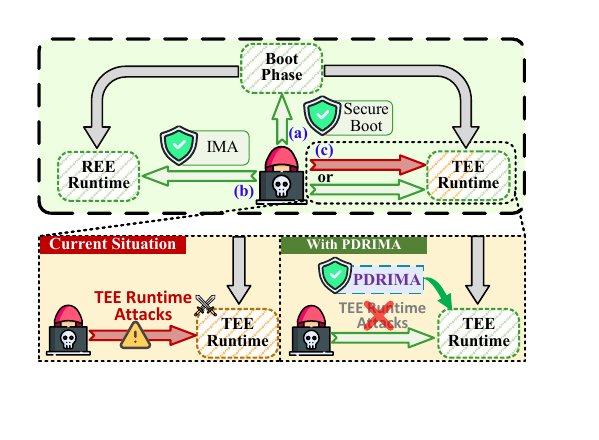} 
\caption{Security coverage of a TrustZone-based device without (bottom-left) and with PDRIMA (bottom-right). 
Existing secure boot and REE-side IMA protect the boot phase and REE runtime (arrows (a), (b)), 
but leave the TEE runtime exposed (arrow (c), bottom-left). PDRIMA closes this gap with
TEE runtime integrity protection (bottom-right).}
\label{fig:intro}
\vspace{-10pt}
\end{figure}

To overcome these limitations, Trusted Execution Environments (TEEs) provide a hardware-backed Root of Trust (RoT) and an isolated environment for critical code and data~\cite{globalplatform_tee_arch}. ARM TrustZone is the dominant TEE technology in embedded and IoT devices, partitioning the platform into a Rich Execution Environment (REE) for normal applications and a TEE for security-sensitive services~\cite{armtrustzonearmv8}. However, due to design and implementation constraints, TrustZone alone does not provide complete secure-boot, runtime integrity protection, or Remote Attestation (RA) mechanisms~\cite{schneider2022sok}, so relying solely on TrustZone cannot guarantee device security. As illustrated by arrows (a)–(c) in Fig.~\ref{fig:intro}, TrustZone-based devices remain vulnerable to both boot-time and runtime attacks~\cite{cerdeira2020sok, koutroumpouchos2021building, de2021toctou, chen2017downgrade, machiry2017boomerang, busch2024spill}. Therefore, achieving end-to-end trust for TrustZone-enabled devices requires a full life-cycle protection mechanism that combines (i) a \textbf{secure boot mechanism} and (ii) \textbf{runtime integrity protection} that continuously verifies the executing software stack on the device~\cite{ling2021secure}.

\textbf{For secure boot mechanism}, existing works~\cite{ling2021secure, lazarou2025measured, wang2024towards, mao2025tpm2} build secure and trusted chains anchored in a hardware-based RoT. As illustrated in Fig.~\ref{fig:intro}, these schemes verify boot-able components before execution to effectively mitigate boot-time attacks (arrow~(a)), but the trust chain stops after static verification. They do not track runtime kernel or application state, leaving devices exposed to long-lived runtime attacks beyond boot (arrows~(b) and~(c))~\cite{cerdeira2020sok, machiry2017boomerang, busch2024spill}.

\textbf{For runtime integrity protection}, the Linux Integrity Measurement Architecture (IMA) provides a basic framework for runtime integrity protection~\cite{sailer2004design}, and later work strengthens functions and security of IMA~\cite{ima_appraisal2012, davi2009dynamic, sun2018security, luo2019container, sassu2018simplera, eckel2021userspace}. Recently, many schemes dedicated to move monitoring and protection modules into TEE to harden REE runtime integrity~\cite{kanonov2016secure, ge2014sprobes, dong2020kims, ling2021secure, song2022tz, liu2023tzeamm, song2024dimac}. As illustrated in Fig.~\ref{fig:intro}, these designs mainly target REE-side attacks (arrow~(b)) and provide no visibility into TEE runtime state (arrow~(c)).

In fact, runtime integrity protection must cover both REE and TEE, since the TEE itself has been shown vulnerable through attacks on Trusted Applications (TAs) and TEE kernels~\cite{cerdeira2020sok, de2021toctou, chen2017downgrade, machiry2017boomerang, busch2024spill}. Nevertheless, runtime integrity protection for TrustZone-based TEEs remains limited. To date, the only deployed mechanism is OPTEE-RA~\cite{suzaki2025openssf}, which measures and attests user-TA inside the TEE. However, it offers narrow coverage and limited functionality for TEE runtime protection, so attacks can still bypass it and compromise TEE integrity~\cite{chen2017downgrade} (detailed in Section~\ref{sec:threat_model}). Therefore, there is a strong need for a more comprehensive TEE runtime integrity solution. However, achieving such protection still faces at least \textbf{three critical challenges}:
\begin{itemize}
    \item \textbf{C1: Vulnerable to some attacks}.
    
    \item \textbf{C2: Lack of Suitable TEE Runtime Protection}.

    \item \textbf{C3: Limited RA for TEE Runtime State}.
\end{itemize}

To address these challenges, we propose \emph{P}olicy-\emph{D}riven \emph{R}untime \emph{I}ntegrity \emph{M}easurement and \emph{A}ttestation (\textbf{PDRIMA}), a novel approach that \textbf{fills the remaining gap in full life-cycle security and trust for TrustZone-based devices}, as illustrated by arrow~(c) in Fig.~\ref{fig:intro} (bottom-right). PDRIMA first systematically analyzes the attack surfaces that affect the runtime integrity of TrustZone-based TEEs, and then designs two subsystems, the Secure Monitor Agent (SMA) and the Remote Attestation Agent (RAA), to provide policy-driven TEE runtime integrity solution across these surfaces and to support on-demand RA of TEE state. Our approach is orthogonal to existing secure boot schemes~\cite{ling2021secure, lazarou2025measured, wang2024towards, mao2025tpm2} and REE runtime integrity mechanisms~\cite{kanonov2016secure, ge2014sprobes, dong2020kims, ling2021secure, song2022tz, liu2023tzeamm, song2024dimac}, and can be deployed alongside them without interference, enabling protection over the device full life-cycle. \textbf{To the best of our knowledge, this is the first work that implements runtime integrity measurement and RA with full coverage of the attack surfaces that may compromise the runtime integrity of TrustZone-based TEEs.}

The contributions of this paper are summarized as follows:
\begin{itemize}
    \item \textbf{Comprehensive TEE Attack Surface Analysis}. We systematically analyze the runtime attack surfaces that affect TrustZone-based TEEs, using OP-TEE as a representative case. Thus, PDRIMA addresses \textbf{C1}.
    
    \item \textbf{Policy-Driven TEE Runtime Integrity Solution Design and Implementation}. We design the SMA subsystem in PDRIMA to provide comprehensive runtime integrity protection for TrustZone-based TEEs across the three identified attack surfaces, thus addressing \textbf{C2}.
    
    \item \textbf{RA Design and Implementation}. We design the RAA subsystem to export tamper-evident integrity evidence from SMA and support RA of the TEE’s runtime integrity state, thus addressing \textbf{C3}.
\end{itemize}

 We implemented a prototype on the Raspberry Pi 3B+ to demonstrate the feasibility of our design and evaluate its performance, showing that PDRIMA is practical in resource-constrained environments.

The rest of this paper is organized as follows. Section~\ref{sec:background_related} introduces the background and related work. Section~\ref{sec:system} presents the threat model, security goals, and an overview of the PDRIMA system. Sections~\ref{sec:design} and~\ref{sec:implementation} describe the design and implementation of PDRIMA, respectively. Section~\ref{sec:evaluation} provides the security and performance evaluation. Finally, Section~\ref{sec:conclusion} concludes the paper and outlines future work.

\section{Background and Related Work}
\label{sec:background_related}

This section describes the background to PDRIMA in Section~\ref{sec:background}. Section~\ref{sec:related} reviews related work on runtime integrity measurement.

\subsection{Background}
\label{sec:background}

This section presents the background, including the ARM TrustZone, OP-TEE, and the principles of integrity measurement.

\subsubsection{ARM TrustZone and OP-TEE}

This section introduces ARM TrustZone, a TEE technology commonly used in embedded and IoT devices; and OP-TEE, a popular open-source TEE Operation System (OS) implementation based on ARM TrustZone-based TEE.

\textbf{ARM TrustZone}. ARM TrustZone provides hardware isolation by partitioning the platform into the Normal World (NW, or REE) and the Secure World (SW, or TEE)~\cite{pinto2019demystifying}. Isolation is enforced by the Non-Secure (NS) bit and by memory/peripheral access control. ARMv8-A defines four Exception Levels (EL0–EL3). The Secure Monitor, such as ARM Trusted Firmware (ATF), runs at EL3 and switches NW/SW on a Secure Monitor Call (SMC) by saving context, toggling NS, and restoring the target world~\cite{armtrustzonearmv8}. Both worlds maintain separate privilege levels: EL0 and EL1 in REE run Client Applications (CAs) and the REE kernel, such as Linux kernel, while S-EL0 and S-EL1 in TEE run Trusted Applications (TAs) and the TEE kernel. Secure boot establishes initial trust by verifying REE/TEE kernel images before execution.

\textbf{OP-TEE}. Open Portable Trusted Execution Environment (OP-TEE) is an open-source TEE OS implementation that follows GlobalPlatform (GP) specifications~\cite{pinto2019trusted}. The OP-TEE core at S-EL1 manages secure memory, threads, and crypto. Pseudo-TAs (pTAs) are statically linked services with kernel privilege, while user-TAs are ELFs loaded on demand by \texttt{ldelf}, which are identified by Universally Unique IDentifiers (UUIDs). OP-TEE provides \textit{TEE Client API} and \textit{TEE Internal Core API} for developing CAs and TAs. A user-space daemon, \textit{tee-supplicant}, runs in the REE to provide user-TA loading and file-system access on behalf of the TEE and to read/write shared memory. Our work builds on these architectural elements and targets runtime integrity in OP-TEE without changing GP semantics or the world-switch path.

\subsubsection{Principles of Integrity Measurement}

This section introduces Integrity measurement mechanism commonly used for runtime integrity protection. This is a effective method for verifying that system remains in a known-good state. The IMA introduced in Linux version 2.6.30 implements it by hashing content on access and maintaining an log of measurements~\cite{sailer2004design}. IMA interposes on file use (e.g., execute, read), computes a cryptographic digest, and extends it to an tamper-evident list. With a Trusted Platform Module (TPM), digests can be extended into Platform Configuration Registers (PCRs) to produce a succinct state summary for verification~\cite{tcg2016tpm}.

IMA provides four core functions~\cite{ima_sourceforge}: \emph{collect} (compute hashes), \emph{store} (maintain a protected list and PCR extensions), \emph{appraise} (compare against trusted “golden” values, often in extended attributes), and \emph{attest} (sign logs for remote verifiers using TPM-protected keys). A policy controls what to measure and when, using rules keyed on object type, path, owner, and operation. The measurement list forms a tamper-evident sequence and supports both local appraisal and RA.

These principles motivate a TEE-oriented design: selective, policy-driven collection; protected, append-only logging; aggregation into registers for efficient checks; and a signed report for RA.

\subsection{Related Work}
\label{sec:related}

This section reviews and outlines the evolution of related work on runtime integrity measurement. We then compare PDRIMA with several representative schemes.

\subsubsection{Traditional Measurements for REE}

Traditional mechanisms lay the foundation for policy-driven measurement and tamper-evident evidence in REE kernels. The representative and foundational work was the IMA~\cite{sailer2004design} proposed in 2004. IMA-Appraisal~\cite{ima_appraisal2012} enhanced IMA with local appraisal against pre-computed golden values, reducing the risk of offline tampering and rollback attacks.

Beyond these baselines, many works extended IMA in different directions. Davi~\emph{et al.} proposed DynIMA~\cite{davi2009dynamic}, which dynamically adjusts IMA to improve security. Sun~\emph{et al.}~\cite{sun2018security} and Lou~\emph{et al.} ~\cite{luo2019container} adjusted the coverage of IMA and split measurement logs and PCRs to support container–aware IMA. Sassu~\emph{et al.}~\cite{sassu2018simplera} design a protocol that refines IMA-based RA by sealing keys to predictable IMA PCR values. Eckel~\emph{et al.}~\cite{eckel2021userspace} proposed USIM, which measures user-space text files that are not covered by IMA. Although these improvements strengthen IMA, they still depend on a correct REE kernel and thus inherit its attack surface, allowing attackers to undermine its security guarantees.

\subsubsection{TEE-based Measurements for REE}

To reduce reliance on REE, many works deploy the monitor inside a TEE. Deyannis~\emph{et al.} proposed SGX-Mon~\cite{deyannis2020enclave}, which places monitor inside an SGX-based enclave to verify the state of an untrusted OS. Galanou~\emph{et al.}~\cite{galanou2023trustworthy} proposed Revelio for applications integrity verification inside Confidential Virtual Machines (CVMs). Other studies~\cite{pecholt2022cocotpm, mao2025towards, narayanan2023remote} apply virtual TPMs to CVMs so that integrity evidence can be recorded and attested without trusting the underlying hypervisor. In addition, several studies focus on heterogeneous devices, such as FPGA platform, using TEE technologies as the RoT to protect its integrity~\cite{wang2024towards, mao2025tpm2}.

On ARM TrustZone, researches deploy monitors in SW for monitoring the REE. Samsung TIMA (Knox)~\cite{kanonov2016secure} performs periodic kernel checks and protects sensitive kernel structures from S-EL1, providing supervision of the REE kernel (Android). SPROBES~\cite{ge2014sprobes} instruments sensitive kernel instructions, enforcing kernel code integrity. KIMS~\cite{dong2020kims} computes and compares digests of the REE kernel and dynamic data structures under TEE control. Ling~\emph{et al.}~\cite{ling2021secure} proposed a paging-based process integrity measurement for TrustZone-based IoT devices, enabling monitoring of REE runtime state at the process granularity. TZ-IMA~\cite{song2022tz} relocates IMA’s measurement and appraisal modules into the TEE and encrypts the measurement log, thereby enhancing IMA with TrustZone protection; it also defines a RA protocol that allows verifiers to verify and assess the integrity evidence of REE. TZEAMM~\cite{liu2023tzeamm}  builds a measurement and fallback mechanism for REE applications based on TrustZone, and Dimac~\cite{song2024dimac} monitors container life-cycle events and records measurements in container-specific logs and PCRs, providing a RA protocol for container integrity.

\subsubsection{TEE-based Measurements for TEE}

Recently, many security issues and exploitable vulnerabilities have been reported in TEEs~\cite{machiry2017boomerang, cerdeira2020sok, busch2024spill}. A compromised TEE OS or privileged TA can break the trust assumptions of REE-based monitors and other security services that rely on the TEE. Consequently, recent work has begun to move runtime integrity measurement into the TEE itself and to cover the runtime state of trusted kernels and applications.

Several solutions start to measure the runtime state of TEEs. Zhou~\emph{et al.}~\cite{zhou2022smile} proposed SMILE, which allows enclave owners to check the integrity of SGX enclaves. Morbitzer~\emph{et al.}~\cite{morbitzer2023guarantee} presented GuaranTEE, which uses control-flow attestation to ensure the integrity of a service running within a TEE. Ozga~\emph{et al.}~\cite{ozga2021triglav} designed TRIGLAV, which attests the runtime integrity of CVMs by binding VM state to a secure login channel.

Within TrustZone-based devices, there is still limited research on runtime integrity measurement. To date, the only implemented solution is OPTEE-RA, proposed by Suzaki~\emph{et al.}~\cite{suzaki2025openssf}. OPTEE-RA implements a pTA for measuring and RA. While OPTEE-RA provides an first solution toward TEE-side monitoring, its monitoring scope and functionality are limited to individual TA binaries and coarse-grained evidence, making it difficult to comprehensively demonstrate the TEE’s runtime integrity status.

\section{System Overview}
\label{sec:system}

This section details the threat model, security goals and the system description of PDRIMA.

\begin{figure*}[!htbp]
    \centering
    \includegraphics[width=1\textwidth,trim=15 20 22 18,clip]{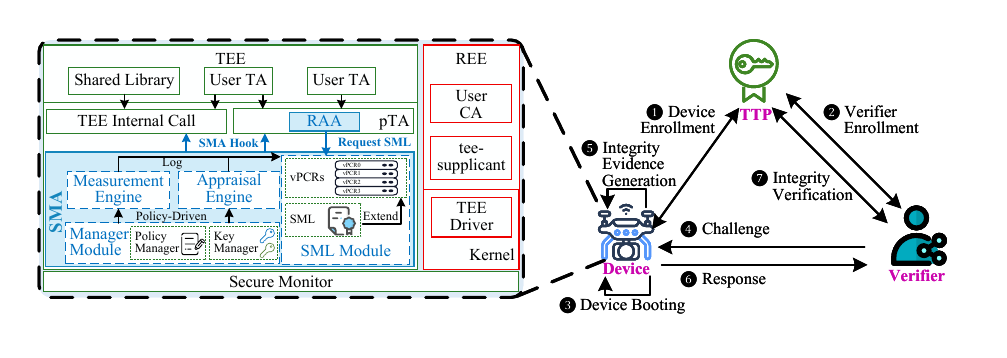} 
    \caption{The architecture and the workflow of the PDRIMA. The left side of the figure depicts the PDRIMA architecture, with the SMA and RAA subsystems located within the TEE and comprising five components. The right side of the figure illustrates the PDRIMA workflow, involving three participants and six key steps.}
    \label{fig:desciption}
\end{figure*}

\subsection{Threat Model}
\label{sec:threat_model}
We first introduce adversary capabilities and security assumptions, then analyze three major attack surfaces to establish a threat model.

\textbf{Adversary Capabilities}. We consider a powerful adversary with the following capabilities:
\begin{itemize}
    \item \textbf{REE System Compromise}. The adversary fully controls the REE system, enabling arbitrary code execution. It can also read/write all non-secure memory and Compromise the \textit{tee-supplicant} daemon. Moreover, it can tamper with shared-memory parameters, and corrupt responses from the TEE.
    \item \textbf{REE File system Control}. The adversary can replace TA binaries (ELFs) with downgraded, vulnerable versions. It can also read, replace, or delete persistent storage of the TEE.
    \item \textbf{Network Control}. The adversary can control the channel between the device and remote verifiers. It can intercept, modify, replay, or drop packets in the network channel.
\end{itemize}

\textbf{Security Assumptions}. We assume all hardware and the cryptographic primitives are correctly implemented. Then, we do not consider the physical attacks, side-channel attacks, denial-of-service attack, and the TOCTOU attacks~\cite{de2021toctou}. Finally, we assume secure boot has verified the authenticity and integrity of the ATF and TEE image at boot time\cite{ling2021secure}. 

\textbf{Attack Surfaces Analysis}. Given the above adversary capabilities and security assumptions, we analyzed the three key Attack Surfaces (\textbf{AS}s):
\begin{itemize}
    \item \textbf{AS1: Boot-Time to Runtime Gap}. 
    
    \item \textbf{AS2: REE-TEE Interface Boundary}. 
    
    \item \textbf{AS3: User–Kernel Spaces Privilege Boundary}.
\end{itemize}

\subsection{Security Goals}
Based on the threat model, we establish the following Security Goals/Requirements (\textbf{SR}s) for PDRIMA.
\begin{itemize}
    \item \textbf{SR1: Trusted Deployment}. 
    
    \item \textbf{SR2: Policy-Driven Flexibility}. 
    
    \item \textbf{SR3: AS1 Coverage}.
    
    \item \textbf{SR4: AS2 Coverage}.
    
    \item \textbf{SR5: AS3 Coverage}.
    
    \item \textbf{SR6: Remote Attestability}.
\end{itemize}

\subsection{System Description}
To satisfy \textbf{SR1}-\textbf{SR6}, we propose \textbf{PDRIMA}, a policy-driven runtime integrity measurement and attestation approach for ARM TrustZone-enabled devices. PDRIMA comprises two subsystems: \textbf{Secure Monitor Agent (SMA)} and \textbf{Remote Attestation Agent (RAA)}. Guided by \emph{policies}, the SMA enables runtime integrity protection through \emph{measurement}, \emph{appraisal}, and \emph{logging}. It hooks events that affect integrity and records evidence in the Security Measurement Log (SML). The RAA then requests the SML from the SMA, prepares attestation data, and interacts with remote verifiers. The system and architecture of PDRIMA are illustrated as Fig.~\ref{fig:desciption}, which consists of three participants and five components.

\textbf{Participants}. Denoted by purple letters in Fig.~\ref{fig:desciption} (right).
\begin{itemize}
    \item \textbf{Trusted Third Party (TTP)}. The TTP is the enrollment authority. It provisions the device and verifier with identities and certificates. It delivers user-TA binaries and customized TEE kernel images that embed PDRIMA. It also maintains reference hash values (“golden hashes”) for the kernel and TAs.
    
    \item \textbf{Device}. The device is the TrustZone-based platform requiring TEE runtime integrity verification. We assume it can securely boot the TEE and support running PDRIMA. It supports networking and can establish secure channels to the verifier.
    
    \item \textbf{Verifier}. The verifier is the remote challenger. It can set up a secure session with the device and generate a fresh nonce. It does not store reference hashes; instead, it forwards the received evidence to the TTP and accepts the TTP’s decision.
\end{itemize}

\textbf{Components}. As indicated by the blue dashed box in Fig.~\ref{fig:desciption} (left), the SMA comprises four components, whereas the RAA consists of a single component.
\begin{itemize}
    \item \textbf{SMA: Manager Module}. It manages SMA workflow. It includes: (i) a \emph{Policy Manager} submodule; (ii) a \emph{Key Manager} submodule; (iii) a \emph{Main Control Logic}.
    
    \item \textbf{SMA: Measurement Engine (ME)}. It calculates hashes over objects according to the policies.
    
    \item \textbf{SMA: Appraisal Engine (AE)}. It matches UUIDs in the Reference Measurement List (RML) entries based on policy configuration and compares the digest to classify the object as trusted or untrusted.

    \item \textbf{SMA: SML Module (SMLM)}. It maintains the SML as an append-only, hash-chained log. It also maintains four vPCRs.

    \item \textbf{RAA: RA-pTA}. It is implemented as a pTA inside the TEE core.
\end{itemize}

\textbf{PDRIMA Workflow}. The complete system operation and RA workflow is illustrated in Fig.~\ref{fig:desciption} (right):

(1) \emph{Enrollment} (Step \circnum{1}-\circnum{2}). The TTP provisions the device and verifier with certificates, and delivers a policy-embedded TEE kernel and a TTP-signed RML to the device.

(2) \emph{Secure boot and monitoring} (Step \circnum{3}). The device securely boots the TEE. The SMA loads policies and RML, drives the ME and AE, which collect evidence at runtime and append it to the SML.

(3) \emph{Challenge-Response} (Step \circnum{4}-\circnum{6}). The verifier sends a fresh nonce. The RA-pTA gets SML (and vPCRs), binds them to the nonce, seals/signs the evidence, and returns it.

(4) \emph{Verification} (Step \circnum{7}). The verifier forwards the response to the TTP, which validates freshness and digests against the RML and issues the integrity verdict.

The above workflow provides a comprehensive overview of the device's TEE runtime integrity status.

\section{System Design}
\label{sec:design}

This section presents the design of PDRIMA’s core functions and explains how they interact to protect TEE runtime integrity.

\subsection{Policy Manager Design}
The policy manager is a rule-based controller for SMA actions. Each Policy Rule (PR) is a triple with an action, an event type, and an optional set of conditions:
\begin{equation}\label{eq:pr}
\mathrm{PR}=\langle \text{action}, \text{event}, \langle\text{condition}_1,\text{condition}_2,\ldots\rangle\rangle .
\end{equation}

The action and event are mandatory; the condition set is optional and variable-length. The action states whether to measure or appraise. The event selects the target events. The conditions limit when a rule applies. A policy set contains one or more PRs. During image building, the set is compiled into a binary blob and embedded into the TEE kernel image by the TTP. Thus, changing policy requires rebuilding and re-signing the kernel, which prevents tampering.

At boot, the SMA loads the policy set from the image, parses it, and stores it in secure memory belonging to the TEE kernel. When an event occurs, SMA scans rules in order and stops at the first match. If no rule matches, the SMA takes the default action to bypass measurement/appraisal for that event, limiting cost on non-critical paths.

\subsection{Measurement Mechanism}
The measurement mechanism generates digests of security-critical objects and events. The ME can perform three types of measurements, including: static measurements, dynamic measurements, and time-based re-measurements.

\subsubsection{Static Measurement}
Static measurements occur during system initialization and component loading. It binds measurement to the code that will execute by hashing selected memory segments. This establishes a trusted baseline before any execution and enables detection of unauthorized changes to binaries and initialization data.

\textbf{TEE Kernel Measurement}. When \texttt{ATF} loads the TEE kernel, the ME hashes security-critical segments. It also includes kernel metadata such as function tables and ARM exception tables.

\textbf{Static Components Measurement}. Since pTAs are statically linked into the kernel image, their code and read-only data are already covered by the kernel measurement.

\textbf{User-TA Measurement}. When OP-TEE loads a TA’s ELF, the ME hashes the TA’s some segments before its first execution. It can also measure TA properties if the policy permits.

The measurement process employs a segmented hashing approach. The ME processes each segment independently and calculates its hash. Subsequently, it combines each value into a hash chain to compute the final hash as the result:
\begin{equation}\label{eq:mr}
\mathrm{Result_{segment}}=H\bigl(H(...\bigl(H(0\parallel S_1)\parallel S_2\bigr)\parallel ...\bigr)\parallel S_n)
\end{equation}
where $S_i$ represents the i-th segment. After that, the engine determines whether appraisal is required based on the policy configuration (detailed in Section \ref{sec:sub_appraisal}). Meanwhile, the measurement results are recorded in a new SML entry in SMLM. The hash value of the SML entry also extends to the appropriate vPCRs by event type, enabling verifiers to confirm the trustworthiness of SML (detailed in Section \ref{sec:sub_sml}).

\subsubsection{Dynamic Measurement}
\begin{figure}[!t]
\includegraphics[width=3.5in,trim=20 20 18 15,clip]{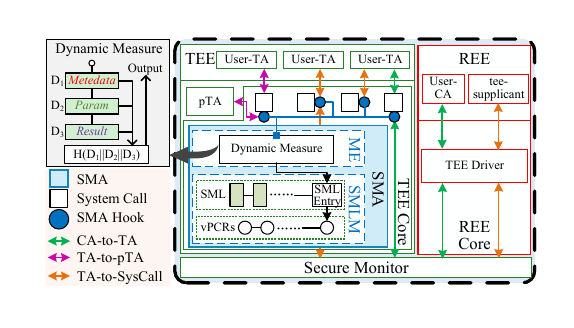} 
\caption{Dynamic measurement mechanism of PDRIMA.}
\label{fig:dynamic_measure}
\vspace{-10pt}
\end{figure}

Dynamic measurement records runtime events including user-TA invocations, inter-TA communications, and security-critical system calls, as illustrated in Fig.~\ref{fig:dynamic_measure}. The SMA installs hooks (SMA Hooks) at selected system call entry points.

\textbf{TA Invocation Measurement}. When a CA in the REE opens a user-TA session or invokes a user-TA command, a hook intercepts the call in the TEE core. The hook records the some parameters and the return code. These records help detect unauthorized access and abnormal invocation patterns.

\textbf{Inter-TA Communication Measurement}. When a CA or user-TA invokes a pTA, the SMA hook at the pTA-related system calls applies the same process.

\textbf{System Call Measurement}. For security-critical system calls, the SMA Hook captures relevant parameters and execution result, proving that a sensitive system call event occurs. 

After interception, the SMA hooks extract identifiers and parameters. The ME then normalizes and serializes this data and computes a digest, as shown in the “Dynamic Measure” box in Fig.~\ref{fig:dynamic_measure}. Each measurement comprises three parts: (i) metadata $D_1$; (ii) parameters $D_2$; and (iii) results $D_3$. The ME binds these parts into a system call measurement result:
\begin{equation}\label{eq:scm}
\mathrm{Result_{syscall}}=H\bigl(D_1\parallel D_2\parallel D_3).
\end{equation}
This composite hash captures the complete operation context. Finally, the ME extends the result as a new entry to the SML in SMLM and extends to vPCRs.

\subsubsection{Time-based Re-Measurement}
\label{sec:sub_remeasure}
To address the integrity verification of long-running components, the SMA supports configurable re-measurement intervals for kernel structures and user-TAs. This mechanism detects runtime modifications that might occur through memory corruption or exploitation.

Formally,

\begin{equation}
\label{eq:recheck}
\text{Re-Measure} \iff T_{\mathrm{current}}-T_{\mathrm{last}}>\textit{Interval}.
\end{equation}
Re-measurement hashes only static segments. The ME then compares the new digest against the baseline recorded during the static measurement process to detect drift or tampering. 

If the comparison succeeds, dynamic measurement proceeds as usual for the intercepted call and restarts the timer; if it fails, the SMA blocks the call or emits an alert, according to policy. In all cases, the re-measurement events are extended as a new entry to the SML, creating an audit trail of system state evolution.

\subsection{Appraisal Mechanism}
\label{sec:sub_appraisal}
The AE determines component integrity by checking measurements against known-good references. It maintains an RML composed of one or more RML Entries (REs). Each RE is a triple with a component UUID, its expected digest, and a minimum acceptable version:
\begin{equation}\label{eq:re}
\mathrm{RE}=\langle \text{UUID}, \text{golden hash}, \text{min version} \rangle .
\end{equation}

During the build process, the TTP constructs the RML and signs it with a dedicated private key $SK_{rml}$. The signature covers the entire RML structure, preventing modification of entries. The TTP then embeds the signed RML and the corresponding public key $PK_{rml}$ into the TEE kernel image. At SMA initialization, the Manager Module loads the $PK_{rml}$, verifies the signature, and only then places a valid RML into secure memory for appraisal operations.

After the ME computes a component hash, the Policy Module specifies whether appraisal is required. If appraisal is enabled, the AE locates an RE by UUID and compares the measured digest with the golden hash. The comparison must match exactly for appraisal to succeed. The AE also enforces version constraints by rejecting components whose version is below the RML-specified minimum, thereby preventing rollback attacks.

\subsection{Evidence Logging Mechanism}
\label{sec:sub_sml}
The evidence logging mechanism in PDRIMA primarily includes extension of the SML and vPCRs.

\begin{figure}[!t]
\includegraphics[width=3.5in,trim=15 15 15 15,clip]{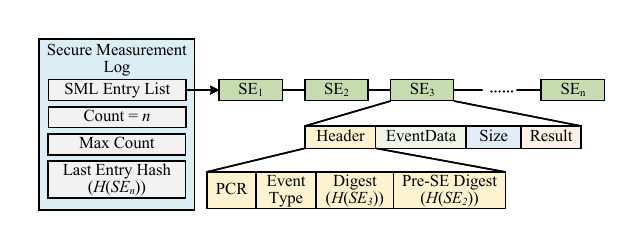} 
\caption{Design of secure measurement log in PDRIMA. SEs are organized in a hash-chained list, where each SE header includes the target vPCR index, event type, its own digest, and the previous SE digest.}
\label{fig:sml}
\vspace{-10pt}
\end{figure}

\textbf{Security Measurement Log (SML)}. The SML is a tamper-evident log structure for storing measurement records. It consists of global metadata and a hash-chained \emph{SML Entry List}, as illustrated in Fig.~\ref{fig:sml}. Each SML Entry (SE) is a quadruple:
\begin{equation}\label{eq:se}
\mathrm{SE}=\langle \text{SEHeader}, \text{event data}, \text{size}, \text{result} \rangle .
\end{equation}
The \textit{SEHeader} contains key metadata for this SE:
\begin{equation}\label{eq:seheader}
\mathrm{SE Header}=\langle \text{vPCR}, \text{event type}, \text{digest}, \text{pre-SE digest} \rangle .
\end{equation}
Here, \textit{digest} authenticates the current SE, and \textit{pre-SE digest} binds it to the previous SE, creating a cryptographic dependency across entries. The \textit{event data} captures event-specific context: for static measurements, measurement results ($Result_{segment}$) of the kernel, pTAs, and user TAs; for dynamic measurements, syscall metadata and parameters ($Result_{syscall}$).

The SML is append-only: a new SE is added to the tail; existing entries cannot be modified or deleted.  This preserves a complete audit trail. The SMA pre-allocates SML storage during initialization to avoid dynamic memory allocation during measurement operations.

\textbf{vPCR Aggregation}. The SMA maintains four vPCRs to summarize the log for attestation: Kernel measurements extend to vPCR[0]; static component measurements to vPCR[1]; user-TA measurements to vPCR[2]; and dynamic measurements to vPCR[3]. Each extension hashes the concatenation of the current vPCR value and the new measurement digest $m$:
\begin{equation}\label{eq:vpcr_extend}
vPCR[i]\ \leftarrow\ H\bigl(vPCR[i]\ \parallel\ m\bigr).
\end{equation}

Verifiers can validate the SML in batch by recomputing chain hashes and comparing the result to the values of vPCRs. They can also perform selective checks over relevant components (e.g., only kernel or only a TA’s entries) to validate specific security properties efficiently.

\subsection{The PDRIMA RA Protocol}
The protocol lets an external verifier assess a device’s TEE runtime integrity. It uses a challenge–response flow with freshness and authenticated evidence, as shown in Fig.~\ref{fig:desciption} (right). 

This workflow consists of 6 steps.

(1) Step \circnum{1}-\circnum{2}. Device and verifier enroll with the TTP. The device can obtain related cryptographic keys such as $PK_{TTP}$, and $\mathrm{Cert}_{TTP}(PK_{Dev})$ for establishing the secure channel with the verifier. Meanwhile, the TEE image embeds $PK_{Attest}/SK_{Attest}$ to sign the integrity evidence. The verifier holds $PK_{Verifier}/SK_{Verifier}$ and $\mathrm{Cert}_{TTP}(PK_{Verifier})$ and can obtain $PK_{TTP}$ to validate certificates.

(2) Step \circnum{3}. The device securely boots. The SMA initializes, loads $PK_{Attest}/SK_{Attest}$, and begins collecting integrity evidence (SML, vPCRs).

(3) Step \circnum{4}. The verifier first establishes a secure channel with the device. Then, it sends a challenge with a fresh $Nonce$ via the REE network stack to the RA-pTA.

(4) Step \circnum{5}. Upon receiving the challenge, the RA-pTA requests Integrity Evidence (IE), including a snapshot of the vPCRs and the SML, from the SMA.
Subsequently, RA-pTA concatenates the $Nonce$ and IE into an Attestation Evidence (AE) and computes a digest. Next, RA-pTA obtains the $SK_{Attest}$ to generate a $Quote$ by signing the AE, which acts as a digital seal guaranteeing the authenticity and integrity of the entire evidence.

(5) Step \circnum{6}. RA-pTA concatenates AE and $Quote$ as the response to the challenge for verification.

(6) Step \circnum{7}. The verifier forwards $AE$ and $Quote$ to the TTP. The TTP (i) verifies $Quote$ with $PK_{Attest}$, (ii) checks $H(AE)$ and the $Nonce$ for integrity and freshness, and (iii) replays the SML against reference values and validates vPCRs to decide device integrity.

\section{Evaluation}
\label{sec:evaluation}

We provide the performance evaluation in this section.

\begin{table}[!t]
\centering
\caption{Comparison of Initialization Times Between PDRIMA and Original OP-TEE}
\label{tab:boot_time}
\renewcommand{\arraystretch}{1.2}
\setlength{\tabcolsep}{4pt}
\begin{tabular}{l c c c}
\toprule
\textbf{Phase} & \textbf{Origin} & \textbf{PDRIMA} & \textbf{Overhead} \\
\toprule
\textbf{Kernel Initialization} & 285 ms & 298 ms & $\approx4.6\%$ \\
\midrule
\textbf{SMA Initialization (1.6 KB; 0.7KB)} & N/A & 39 ms & -  \\
\midrule
\textbf{Kernel Measurement (305 KB)} & N/A & 20 ms & -   \\
\midrule
\textbf{Baseline Generation (284 KB)} & N/A & 15 ms & -   \\
\midrule
\textbf{pTA Measurement (Count: 7)} & N/A  & 4 ms & - \\
\toprule
\textbf{Total Time} & 285 ms & 376 ms & $\approx31.9\%$  \\
\bottomrule
\end{tabular}
\end{table}

\subsection{Performance Evaluation}
\label{sec:sub_performance}

This section presents the performance evaluation of PDRIMA, focusing on initialization and runtime overhead.

\subsubsection{Experiment Setup}

All experiments were run on a Raspberry Pi 3B+ with a quad-core Arm Cortex-A53 (ARMv8) and 1 GB SDRAM. The REE is Raspbian Linux (kernel 4.19.93-v7); the TEE is OP-TEE 3.16.0. All measurements are collected from OP-TEE kernel output to avoid user-space variability.

\subsubsection{Initialization Overhead}

Table~\ref{tab:boot_time} compares original OP-TEE with PDRIMA enabled. The original OP-TEE kernel initialization takes 285~ms. With PDRIMA, total initialization time increases to 376~ms, resulting in an overhead of approximately 31.9\%.

This one-time initialization cost is acceptable for IoT or embedded devices. In exchange, PDRIMA establishes a comprehensive measurement baseline for the kernel and all pTAs at boot time, which the original OP-TEE lacks.

\begin{figure}[!t]
\includegraphics[width=3.5in,trim=5 9 5 5,clip]{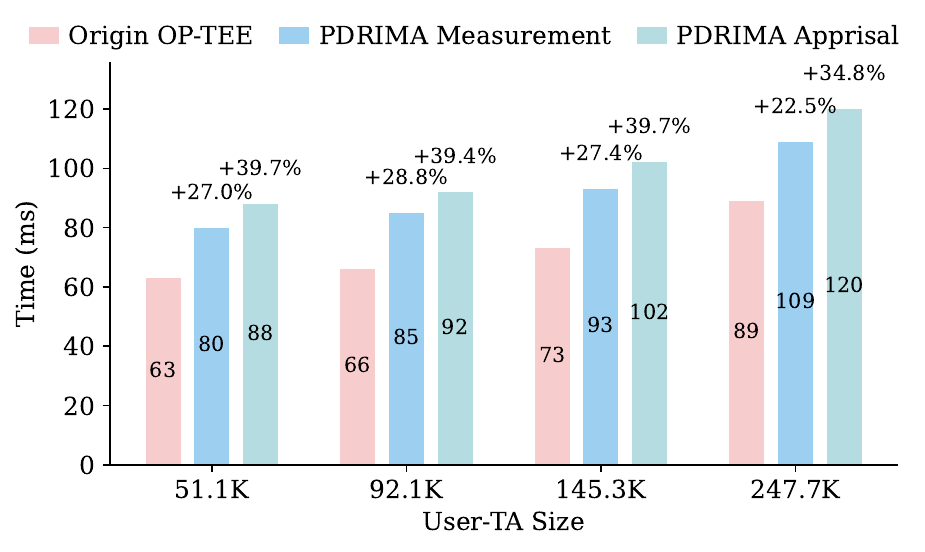} 
\caption{Storage and cryptographic system call latency across varying data sizes and cryptographic parameter configurations under original OP-TEE and PDRIMA.}
\label{fig:ta_load_time}
\vspace{-10pt}
\end{figure}

\subsubsection{Runtime Overhead}

We compare the runtime overhead of original OP-TEE with PDRIMA under three aspects: (i) user-TA loading, (ii) time-based re-measurement, and (iii) interception of security-critical system calls. 

\textbf{User-TA Loading}. Fig.~\ref{fig:ta_load_time} shows the user-TAs loading latency across four TA sizes (51.1~KB-247.7~KB) for three configurations: the original OP-TEE as baseline, PDRIMA with measurement-only mode, and PDRIMA with appraisal mode. For the 51.1 KB TA, load time rises from 63 ms (baseline) to 80 ms (measurement, +27.0\%) and 88 ms (appraisal, +39.7\%). As TA size increases, the absolute loading time grows for all three configurations, while the measurement overhead remains relatively stable at approximately 22–28\%, whereas the appraisal overhead consistently stays around 35–40\%. The overhead stems from two primary sources: (i) hashing TA segments and updating the SML/vPCRs in the measurement phase, and (ii) RML lookup, golden-hash comparison, and version checks in the appraisal phase, which add approximately 8–11~ms. Since most TAs load once, this cost is amortized and acceptable for typical TEE applications.

\section{Conclusion}
\label{sec:conclusion}

In this paper, we presented PDRIMA, a policy-driven runtime integrity measurement and attestation approach for ARM TrustZone-based TEEs. To the best of our knowledge, PDRIMA is the first approach to provide comprehensive TEE runtime integrity protection, filling the remaining gap in the full life-cycle security of TrustZone-based TEE.

\FloatBarrier
\bibliographystyle{jabbrv_IEEEtran}
\bibliography{main}


\end{document}